\documentclass[12pt,preprint]{aastex}

\newcommand{\Hsi}{\textit{Reuven Ramaty High Energy Solar Spectroscopic Imager}}
\newcommand{\hsi}{\textit{RHESSI}}
\newcommand{\goes}{\textit{GOES}}
\newcommand{\sm}{$\sim$}
\newcommand{\Sdo}{\textit{Solar Dynamics Observatory}}
\newcommand{\sdo}{\textit{SDO}}
\newcommand{\hmi}{Helioseismic and Magnetic Imager}
\newcommand{\hinode}{\textit{Hinode}}
\newcommand{\ha}{H$\alpha$}
\newcommand{\aia}{Atmospheric Imaging Assembly}
\def\mathbi#1{\textbf{\em #1}}

\begin{document}

\title{Response of the Photospheric Magnetic Field to the X2.2 Flare on 2011 February 15}
\author{Shuo Wang\altaffilmark{1}, Chang Liu\altaffilmark{1}, Rui Liu\altaffilmark{1}, Na Deng\altaffilmark{1}, Yang Liu\altaffilmark{2}, and Haimin Wang\altaffilmark{1}}
\affil{1. Space Weather Research Laboratory, New Jersey Institute of Technology, University Heights, Newark, NJ 07102-1982, USA; haimin.wang@njit.edu}
\affil{2. W.~W. Hansen Experimental Physics Laboratory, Stanford University, Stanford, CA 94305-4085, USA}

\begin{abstract}
It is well known that the long-term evolution of the photospheric magnetic field plays an important role in building up free energy to power solar eruptions. Observations, despite being controversial, have also revealed a rapid and permanent variation of the photospheric magnetic field in response to the coronal magnetic field restructuring during the eruption. The \hmi\ instrument (HMI) on board the newly launched \Sdo\ (\sdo) produces seeing-free full-disk vector magnetograms at consistently high resolution and high cadence, which finally makes possible an unambiguous and comprehensive study of this important back-reaction process. In this study, we present a near disk-center, \goes-class X2.2 flare, which occurred in NOAA AR 11158 on 2011 February 15. Using the magnetic field measurements made by HMI, we obtained the first solid evidence of a rapid (in about 30 minutes) and irreversible enhancement in the horizontal magnetic field at the flaring magnetic polarity inversion line (PIL) by a magnitude of $\sim$30\%. It is also shown that the photospheric field becomes more sheared and more inclined. This field evolution is unequivocally associated with the flare occurrence in this sigmoidal active region, with the enhancement area located in between the two chromospheric flare ribbons and the initial conjugate hard X-ray footpoints. These results strongly corroborate our previous conjecture that the photospheric magnetic field near the PIL must become more horizontal after eruptions, which could be related to the newly formed low-lying fields resulted from the tether-cutting reconnection.

\end{abstract}


\section{INTRODUCTION}
Almost two decades ago, we discovered rapid and permanent changes of vector magnetic fields
associated with flares \citep{wang92,wang94}. Specifically, the transverse field near the flaring magnetic polarity inversion line (PIL) is found to enhance substantially and irreversibly across the time duration of the flare, which is also often accompanied by an increase of magnetic shear. Similar trend indicating a more horizontal orientation of the photospheric magnetic field after flares and coronal mass ejections (CMEs) has continued to be observed later on in many observations \citep{wang02b,wang04,wang+liu05,liu05,wang07a,jing08,li09,chang11}, and shows some agreement with recent model predictions \citep{li10}. Nevertheless, a majority of such studies are unavoidably hampered by the obvious limitations, ground-based observations (e.g., seeing variation and the limited number of observing spectral positions), probably because of which mixed results were also reported \citep{ambastha93,hagyard99,chen94,li00a,li00b}.

On the other hand, flare-related variations in the line-of-sight (LOS) component of photospheric magnetic field have been clearly recognized \citep[e.g.,][]{wang02b,spirock02,yurchyshyn04,sudol05,wang06,wang10,petrie10}. In particular, the feature of unbalanced flux evolution of the opposite polarities could provide an indirect evidence for the more horizontal orientation of photospheric fields after flares/CMEs \citep{wang10}. However, it is noted that the changes of the LOS field alone cannot provide complete understanding of the field restructuring \citep{hudson11}.

It is notable that vector magnetic field data has been made available with the \hmi\ (HMI) instrument \citep{schou11} on board the newly launched \Sdo\ (\sdo). Its unprecedented observing capabilities give a favorable opportunity to finally resolve any uncertainties regarding the evolution of photospheric magnetic field in relation to flares/CMEs.

In this study, we investigate a near disk-center X2.2 flare on 2011 February 15, which provides the first solid evidence of the enhancement in the horizontal field at the flaring PIL using the seeing-free HMI data. We will discuss the implications of such a change in the context of magnetic reconnection model for flares.

\section{OBSERVATIONS AND DATA REDUCTION}
The HMI instrument obtains filtergrams in six polarization states at six wavelengths along the Fe~{\sc i} 6173~\AA\ spectral line to compute Stokes parameters {\it I}~{\it Q}~{\it U}~{\it V}, which are then reduced with the HMI science data processing pipeline\footnote{\url{http://jsoc.stanford.edu/jsocwiki/VectorMagneticField}} to (1) retrieve the vector magnetic field using the Very Fast Inversion of the Stokes Vector (VFISV) algorithm \citep{borrero10} based on the Milne-Eddington approximation, (2) resolve the 180$^{\circ}$ azimuthal ambiguity using the ``minimum energy'' method \citep{metcalf94,leka09}. As of the time of this writing, only AR11158 processed data have been released by the HMI team \citep{hoeksema11}. For our study, we use the product of vector magnetograms projected and remapped to heliographic coordinates, with a spatial resolution of $\sim$1\arcsec\ and a cadence of 12 minutes.

The temporal and spatial relationship between the change of the photospheric fields and flare energy release can provide important clues concerning the eruption mechanism. The evolution of the flare hard X-ray (HXR) emission was entirely registered by the \Hsi\ \citep[\hsi;][]{lin02}. PIXON images \citep{hurford02} in the 35--100~keV energy range showing the flare footpoints were reconstructed using the front segments of detectors 2--8 with 16--32~s integration time throughout the event. To provide the chromospheric and coronal context, we also used \ha\ images taken by the Solar Optical Telescope \citep[SOT;][]{tsuneta08} on board \hinode, and EUV images made by the \aia\ \citep[AIA;][]{leman11} on board \sdo.

\section{RESULTS} \label{result}
\begin{figure}[!t]
\epsscale{.91}
\plotone{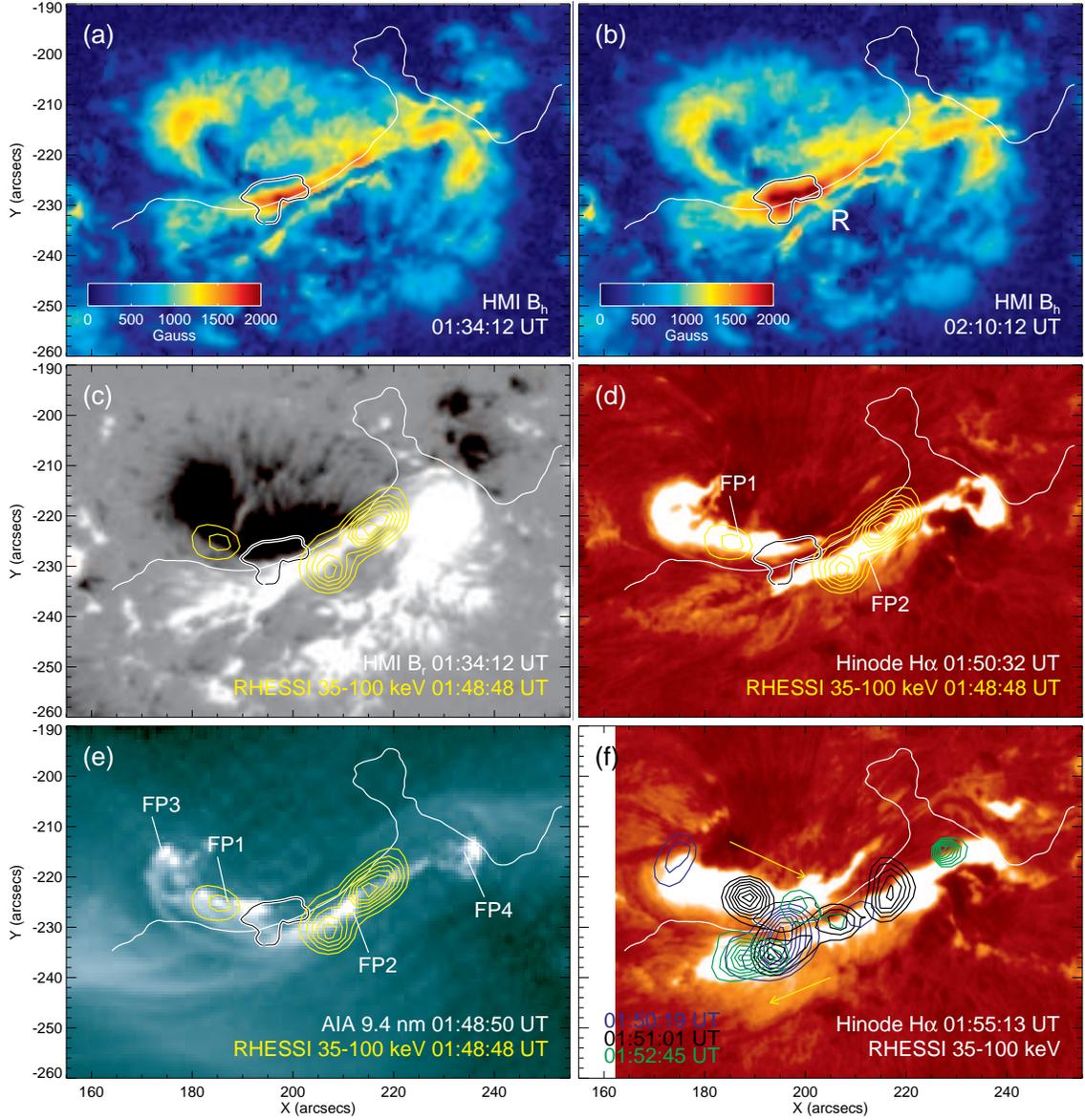}
\caption{Pre- (a) and postflare (b) HMI $B_h$ maps revealing the enhancement of horizontal field in a region R at the PIL (white line) as enclosed by the white bordered line, which is defined based on the smoothed difference image of $B_h$. A preflare $B_v$ map in (c) (scaled at $\pm$1~kG), the first available \hinode/SOT \ha\ image in (d), an AIA 94~\AA\ image at the flare onset in (e), and an \ha\ image at the flare peak time in (f) are overplotted with contours (30\%--90\% of the maximum flux) denoting \hsi\ PIXON images. The arrows in (f) indicate the direction of motion of the main HXR footpoints, as well as that of the chromospheric ribbons besides their separation. \label{f1}}
\end{figure}

\begin{figure}[!t]
\epsscale{.7}
\plotone{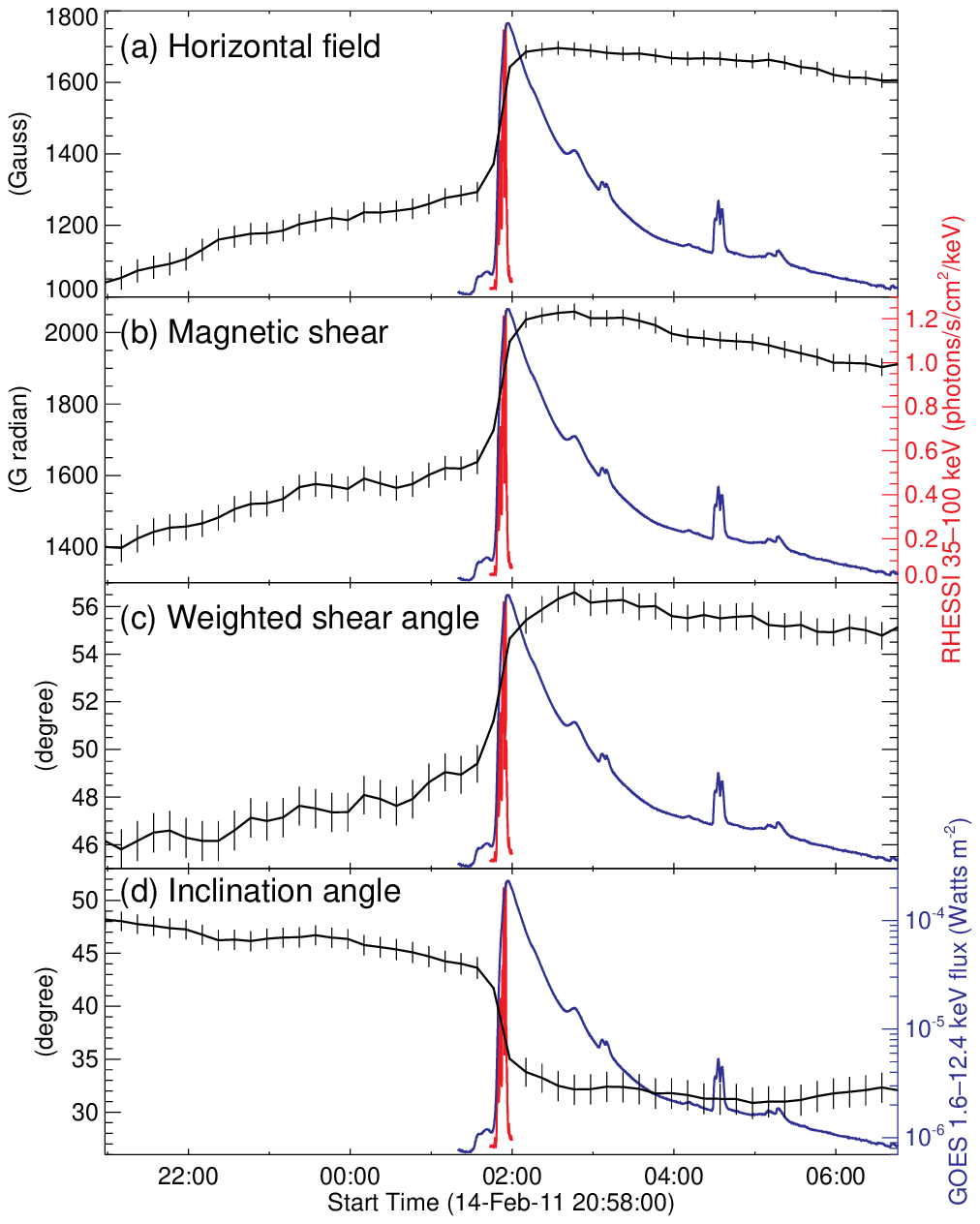}
\caption{Temporal evolution of various magnetic properties of the region R enclosed by the white bordered line in Figure~\ref{f1}, in comparison with the light curves of \hsi\ HXR flux in the 35--100~keV energy range (red) and \goes\ flux in 1--8~\AA\ (blue). The vertical error bars indicate $3\sigma$ level. See \S~\ref{result} for details. \label{f2}}
\end{figure}

The $\beta\gamma$ region NOAA 11158 was located close to the disk center (S21$^{\circ}$, W21$^{\circ}$) when the 2011 February 15 X2.2 flare started at 01:44~UT, peaked at 01:56~UT, and ended at 02:06~UT in \goes\ 1--8~\AA\ flux. The flare was initiated at the center of the active region, where opposite magnetic flux concentrations underwent a counterclockwise rotation-like motion possibly resulting in highly sheared fields along the PIL \citep{liuy11,sun11}. By monitoring the evolution of horizontal field , a compact region R along the PIL (enclosed by the white bordered line) is readily identified to show a pronounced enhancement of horizontal field strength $B_h=\sqrt{B_x^2+B_y^2}$ after the flare (cf. Figures~\ref{f1} (a) and (b)). Close temporal association of this field change with flare emissions and its permanence relative to the flare duration are demonstrated in Figure~\ref{f2}(a) covering a period spanning 10 hrs, in which we find that $\langle B_h \rangle$ at the region R increases by $\sim$30\% from $\sim$1300~G to $\sim$1700~G in \sm30 minutes. This rapid evolution ensues from the beginning of the flare at 01:44~UT, with the change-over time cotemporal with the rapid rising of soft X-ray flux and peaking of HXR emissions. To further characterize the properties of magnetic field, we calculate magnetic shear $\tilde{S}$, weighted shear angle $\mathring{S}$, and magnetic inclination angle $\varphi$. Here $\tilde{S}$ is defined as the product of field strength and shear angle $\tilde{S}=B \cdot \theta$ \citep{wang94,wang06shear,jing08}, where $B=|\mathbi{B}|$, $\theta={\rm cos}^{-1}(\mathbi{B} \cdot \mathbi{B}_p)/(BB_p)$, and the subscript $p$ represents the potential field. The weighted shear angle of a region of interest with n pixels is then $\mathring{S}=\sum_{i} \tilde{S}_i / \sum_i B_i$, where i=1,2,...n. The inclination angle $\varphi$ relative to the horizontal plane is $\varphi = {\rm tan}^{-1} (|B_v| / (B_x^2+B_y^2)^{1/2})$. From the results shown in Figures~\ref{f2}(b)--(d), it can be clearly seen that all of $\langle \tilde{S} \rangle$, $\mathring{S}$, and $\langle\varphi\rangle$ exhibit an abrupt change in the field strength, inclination angle, and azimuthal angle by about 400~G~radian, 7$^{\circ}$, and $-10^{\circ}$, respectively, within the same transition time as $\langle B_h \rangle$ upon the flare occurrence. Please note that in order to demonstrate that the rapid changes are very significant compared to variations seen in the long-term evolution, we plot the $3\sigma_{pre}$ ($3\sigma_{pos}$) as error bars in Figure~\ref{f2}, where $\sigma_{pre}$ ($\sigma_{pos}$) is derived from the linear fit of the temporal evolution of each quantity in the preflare (postflare) state. Corroborating our previous studies \citep[][and references therein]{wang10}, the above rapid developments evidenced by the unambiguous HMI observation strongly suggest a more horizontal and sheared state of the photospheric magnetic field at the region R after the flare. We note that although the increase of $\tilde{S}$ and $\mathring{S}$ seems contrary to the relaxation of nonpotentiality as required to energize eruptions, it has been demonstrated using field extrapolations that the increase is localized and both $\tilde{S}$ and $\mathring{S}$ decrease above a certain height \citep{jing08,sun11,chang11}. The magnetic free energy in the 3-D volume is reduced after the flare.

\begin{figure}[t]
\epsscale{.98}
\plotone{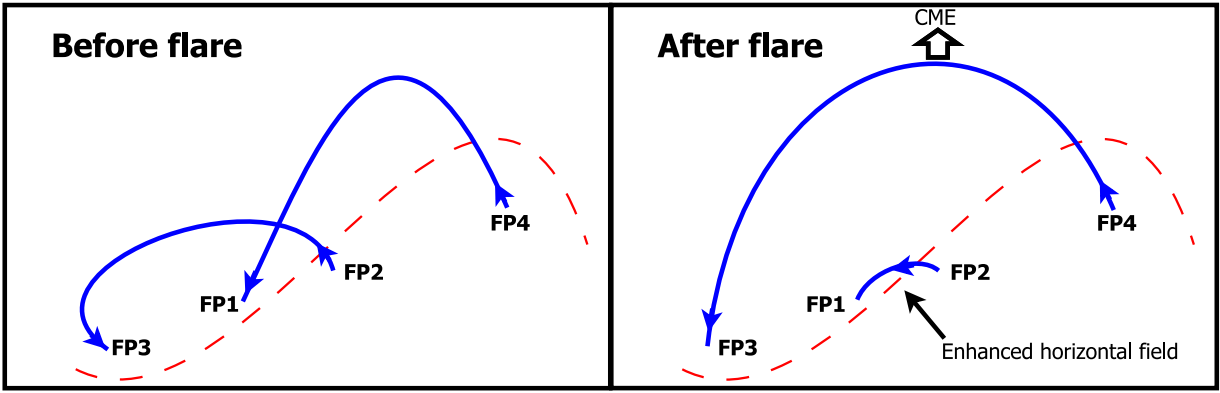}
\caption{Schematic picture interpreting our observations based on the model of \citet{moore01}. Two sigmoidal loops FP3--FP2 and FP4--FP1 in the preflare state (left panel) reconnect to create a large-scale erupting loop FP3--FP4 escaping as a CME and smaller loop FP1--FP2 lying close to the surface contributing to the detected surface magnetic field change (right panel). For clarity, overlying arcade fields and their reconnection leading to flare ribbons are omitted. \label{f3}}
\end{figure}

On the spatial relationship between the field change and flare emissions, the region R lies between the earliest conjugate HXR footpoint sources at \sm01:49~UT (Figure~\ref{f1}(c)) and the ends of the double J-shaped flare ribbons (Figure~\ref{f1}(d)). Intriguingly, AIA 94~\AA\ images show two extra footpoint-like flare brightenings FP3 and FP4 at the two far ends of the flaring PIL (Figure~\ref{f1}(e)). Co-spatial HXR emissions at FP3 and FP4 were observed few minutes later (Figure~\ref{f1}(f)), and the motion of HXR footpoints as well as the evolution of chromospheric ribbons generally proceed in such a manner as to reduce the magnetic shear, along the PIL (illustrated by the arrows in Figure~\ref{f1}(f)) as reported in eruptive sigmoids \citep[e.g.,][]{ji08}. These lead us to a picture as we schematically illustrate in Figure~\ref{f3}, where the flare could be triggered by the tether-cutting reconnection \citep{moore95,moore01} between the two sets of sigmoidal loops FP3--FP2 and FP4--FP1 as clearly discernible in EUV images, which results in the J-shaped flare ribbons \citep[also see][]{schrijver11b}. The reconnected large-scale fields FP3--FP4 could erupt outward to become the halo CME associated with this flare, and the newly formed smaller loops FP1--FP2 lying close to the surface could then account for the enhanced $B_h$ at the region R. Such a reconnection of two current-carrying loops would also effectively lead the current path to move downward closer to the surface, which can explain the increase of $\tilde{S}$ and $\mathring{S}$ \citep{melrose97}. Alternatively, increase of the magnetic nonpotentiality at and near the surface could result from the newly emerging, sheared magnetic flux \citep{jing08}, which could occur after the relaxation of fields above the surface due to the flare energy release. It is worth mentioning that the region R at the PIL is between flare ribbons/kernels at opposite polarities, hence the observed field changes cannot be attributed to flare emissions \citep{patterson81,qiu03}. Detailed investigation of the flare HXR emission in further relation to the coronal field dynamics is out of the scope of the current study and will be presented in a subsequent publication.

\section{SUMMARY AND DISCUSSION}
We have used the unprecedented SDO/HMI vector field observations to analyze the changes of the photospheric magnetic field associated with the first X-class flare in the solar cycle 24, with the aid of images of flare emissions in multiple wavelengths. Main results are as follows.

\begin{enumerate}
\item A compact region R along the flaring PIL shows a rapid and permanent enhancement of $\langle B_h \rangle$ by 400~G (\sm30\% of the preflare magnitude) within about 30~minutes, which has a close temporal relationship with the flare HXR emission. Meanwhile, the nonpotentiality represented by magnetic shear also exhibits a pronounced increase near the surface.

\item The initial HXR sources FP1 and FP2 as well as the double J-shaped flare ribbons are at the two ends of the region R lying at the central of this sigmoidal active region. Two additional flare footpoints FP3 and FP4 are clearly seen in the hot 94~\AA\ channel, located at the far ends of the sigmoid. We suggest that the tether-cutting reconnection \citep{moore01} between the loops FP3--FP2 and FP4--FP1 produces the short and low-lying loops FP1--FP2, which could explain the enhanced $B_h$ as well as $\tilde{S}$ and $\mathring{S}$ at the region R \citep{melrose97}. The detected enhancement of nonpotentiality on the surface could also be due to the newly emerging, sheared fields \citep{jing08}.
\end{enumerate}

In summary, the HMI observations presented in this study constitute the first solid evidence of flare-induced changes of the photospheric magnetic field, which strongly endorses our previous results using ground-based vector magnetograms subject to seeing variation \citep[][and references therein]{wang10}. The unambiguously observed enhancement of horizontal field on the surface strongly suggests that the photospheric magnetic field could respond to the coronal field restructuring by tilting toward the surface (i.e., toward a more horizontal state) near the PIL, and that this development may be due to the tether-cutting reconnection producing the flare. This view is also well in line with the recent theoretical development \citep{hudson08,fisher10}, where the back reaction on the solar surface resulting from the coronal field evolution as required by the energy release is quantitatively assessed. Further systematic studies of flare-related magnetic field change, especially when aided with extrapolation models, are promising to provide further insight into the relationship between the surface field change and coronal magnetic reconnection \citep[e.g.,][]{sun11,chang11}.

\acknowledgments
\sdo\ is a mission for NASA's Living With a Star program. \hsi\ is a NASA Small Explorer. \hinode\ is a Japanese mission developed and launched by ISAS/JAXA, collaborating with NAOJ as a domestic partner, and NASA (USA) and STFC (UK) as international partners. It is operated by these agencies in co-operation with ESA and NSC (Norway). We thank the referee for valuable comments that helped us to improve the paper. S.W., C.L., R.L., and H.W. were supported by NSF grants AGS 08-19662, AGS 08-49453, and AGS 09-36665, and NASA grants NNX 08AQ90G, NNX 08AJ23G, and NNX 11AC05G. N.D. was supported by NASA grant NNX 08AQ32G.

\end{document}